\documentclass[12pt]{article}
\usepackage{graphicx,amssymb,amsmath}
\usepackage{color,graphicx}
\usepackage{cite}
\usepackage{amssymb,amsmath}
\newcommand{\be}{\begin{equation}}
\newcommand{\ee}{\end{equation}}
\newcommand{\bea}{\begin{eqnarray}}
\newcommand{\eea}{\end{eqnarray}}
\newcommand{\beas}{\begin{eqnarray*}}
\newcommand{\eeas}{\end{eqnarray*}}

\begin{document}

\begin{flushright}
 HIP-2014-13/TH
\end{flushright}

\begin{center}

\centerline{\Large {\bf Flowing holographic anyonic superfluid}}

\vspace{8mm}

\renewcommand\thefootnote{\mbox{$\fnsymbol{footnote}$}}
Niko Jokela,${}^{1,2,3}$\footnote{niko.jokela@helsinki.fi}
Gilad Lifschytz,${}^{4}$\footnote{giladl@research.haifa.ac.il} and
Matthew Lippert${}^5$\footnote{M.S.Lippert@uva.nl}

\vspace{4mm}

${}^1${\small \sl Department of Physics} and ${}^2${\small \sl Helsinki Institute of Physics} \\
{\small \sl P.O.Box 64} \\
{\small \sl FIN-00014 University of Helsinki, Finland}

\vspace{2mm}
${}^3${\small \sl Departamento de F\'isica de Part\'iculas} \\
{\small \sl Universidade de Santiago de Compostela}\\
{\small \sl and}\\
{\small \sl Instituto Galego de F\'isica de Altas Enerx\'ias (IGFAE)}\\
{\small \sl E-15782 Santiago de Compostela, Spain} 

\vspace{2mm}
${}^4${\small \sl Department of Mathematics and Physics} \\
{\small \sl University of Haifa at Oranim, Kiryat Tivon 36006, Israel}

\vspace{2mm}
\vskip 0.2cm
${}^5${\small \sl Institute for Theoretical Physics} \\
{\small \sl University of Amsterdam} \\
{\small \sl 1090GL Amsterdam, Netherlands} 

\end{center}

\vspace{8mm}

\setcounter{footnote}{0}
\renewcommand\thefootnote{\mbox{\arabic{footnote}}}

\begin{abstract}
\noindent
 
We investigate the  flow of a strongly coupled anyonic superfluid based on the holographic D3-D7' probe brane model.  
By analyzing the spectrum of fluctuations, we find the critical superfluid velocity, as a function of the temperature, at which the flow stops being dissipationless 
when flowing past a barrier.  We find that at a larger velocity the flow becomes unstable even in the absence of a barrier.
\end{abstract}

\newpage
\tableofcontents

%%%%%%%%%%%%%%%%%%%%%%%%%%%%%%%%%%%%%%%%%%%%%%%%%%%%%%%%%%%%%%%%%%%%%%%%%%%%%%%%%%%%%%%%%%%%%%%%%%%%%%%%%%%%%%%%%%%%%%%%%%%%%%%%%%%%%%%%%%%%%%%%%%%%%%%%%%%%%%%%%%%%%
%%%%%%%%%%%%%%%%%%%%%%%%%%%%%%%%%%%%%%%%%%%%%%%%%%%%%%%%%%%%%%%%%%%%%%%%%%%%%%%%%%%%%%%%%%%%%%%%%%%%%%%%%%%%%%%%%%%%%%%%%%%%%%%%%%%%%%%%%%%%%%%%%%%%%%%%%%%%%%%%%%%%%
%%%%%%%%%%%%%%%%%%%%%%%%%%%%%%%%%%%%%%%%%%%%%%%%%%%%%%%%%%%%%%%%%%%%%%%%%%%%%%%%%%%%%%%%%%%%%%%%%%%%%%%%%%%%%%%%%%%%%%%%%%%%%%%%%%%%%%%%%%%%%%%%%%%%%%%%%%%%%%%%%%%%%
%%%%%%%%%%%%%%%%%%%%%%%%%%%%%%%%%%%%%%%%%%%%%%%%%%%%%%%%%%%%%%%%%%%%%%%%%%%%%%%%%%%%%%%%%%%%%%%%%%%%%%%%%%%%%%%%%%%%%%%%%%%%%%%%%%%%%%%%%%%%%%%%%%%%%%%%%%%%%%%%%%%%%

\section{Introduction}\label{intro}

Recently, a holographic description of anyon superfluidity was presented in \cite{Jokela:2013hta}. The model was based on the D3-D7' system \cite{Bergman:2010gm} with 
a constant magnetic field and charge density in a fractional quantum Hall phase.  By an appropriate $SL(2,{\mathbb{Z}})$
action changing the quantization of the bulk gauge field, this quantum Hall state was transformed into a gapless, anyonic superfluid. In this note, we explore the 
physics of this anyonic superfluid with a nonzero superfluid velocity $v_f$.

%zero temperature, gapless excitation = phonon, Landau criterion 
In a static superfluid, $v_f =0$, the lowest lying excitation is a gapless phonon mode with linear dispersion $\omega = v_s k$. 
For a flowing superfluid at zero temperature, as long as its relative velocity with respect to some barrier or any other object is less than the speed of sound $v_s$, 
the fluid can not exchange energy and momentum with the object; this is the reason the flow is dissipationless.  Above such a critical velocity, energy and momentum can be exchanged, 
and the flow is no longer without dissipation.\footnote{The gapless phonon is necessarily the lowest 
mode at small $k$.  However, at larger $k$, other modes, such as rotons or vortices, can have smaller energy, in which case the critical superfluid velocity is given by the minimum of $\omega/k$.}

This critical velocity can be understood by looking at the excitation spectrum of the flowing superfluid. 
 At zero temperature, a moving superfluid can be obtained simply by a Lorentz boost of a static superfluid.  The excitation spectrum of the flowing superfluid is likewise just 
 the Lorentz transform of the zero-velocity spectrum.  As the superfluid flows faster, the velocity of phonons in the opposite direction decreases.  
 When $v_f > v_s$, antiparallel phonons have negative energy at small momentum, signaling that the energy of the superfluid can be lowered by exciting them.  
If there are objects, such as impurities or the walls of the capillary, that can excite these modes, then 
the flow stops being dissipationless.  This criterion for superfluid stability is called the Landau criterion\cite{Landau}.

%nonzero temperature
At nonzero temperature, Lorentz symmetry is broken, and so the fluctuation spectra must actually be computed.  There is still a critical superfluid velocity $v_{crit}$
above which the gapless phonons develop a negative dispersion and dissipation can occur.  However, in general, this is less than the speed of sound; \emph{i.e.} $v_{crit} < v_s$.

%holographic model
In the holographic model we have an infinite, homogeneous superfluid without any barriers or impurities, but we can still compute the critical velocity by a fluctuation analysis. 
It turns out that there are three important velocity thresholds for the superfluid. First, there is the Landau critical velocity $v_{crit}$, 
which is the velocity of the superfluid above which the energy of the backward directed phonons becomes negative. 
This would signal an instability towards dissipative flow if there were an object which could excite these modes.  However, we also find another, larger velocity above which the phonon dispersion acquires a positive imaginary part,
signaling a spontaneous instability of the flow, even if no object is present. 
We label this $v_{complex}$.  Finally, there is $v_{max}$, the velocity above which the solution of the 
equations of motion ceases to exist. Interestingly, we find that at zero temperature $v_{max}$ is universal and equal to the speed of light, irrespective of the mass of the ambient fermions.
We compute these velocities as functions of the temperature and comment on their physical meanings. 

%comparison with Amado et al
 A similar analysis for a conventional holographic superfluid was performed in \cite{Amado:2013aea}, where it was found that, when the velocity of the phonons became negative, these modes also became tachyonic.  
 In contrast, we find $v_{crit} <  v_{complex}$.  In addition, while the model of \cite{Amado:2013aea} could only be analyzed at relatively high temperature, the special nature of our probe brane 
 model allows trustworthy analysis all the way to zero temperature.  
 
 %organization
This paper is organized as follows. We begin in Sec.~\ref{sec:model} by briefly reviewing the D3-D7' model, the proxy of a fractional quantum Hall state, at nonzero background 
magnetic field and charge density, and in the presence of an electric field. In Sec.~\ref{sec:flowingsuperfluid}, we perform the $SL(2,{\mathbb{Z}})$ transformation to map the 
system to an anyon superfluid phase with zero effective electric and magnetic fields but possessing a nonvanishing current.  We present an analysis of the fluctuations of 
the flowing superfluid in Sec.~\ref{sec:fluctuations} and discuss the physical meaning of the result.  And finally, we summarize our results and discuss open questions in Sec.~\ref{sec:discussion}.

%%%%%%%%%%%%%%%%%%%%%%%%%%%%%%%%%%%%%%%%%%%%%%%%%%%%%%%%%%%%%%%%%%%%%%%%%%%%%%%%%%%%%%%%%%%%%%%%%%%%%%%%%%%%%%%%%%%%%%%%%%%%%%%%%%%%%%%%%%%%%%%%%%%%%%%%%%%%%%%%%%%%%
%%%%%%%%%%%%%%%%%%%%%%%%%%%%%%%%%%%%%%%%%%%%%%%%%%%%%%%%%%%%%%%%%%%%%%%%%%%%%%%%%%%%%%%%%%%%%%%%%%%%%%%%%%%%%%%%%%%%%%%%%%%%%%%%%%%%%%%%%%%%%%%%%%%%%%%%%%%%%%%%%%%%%
%%%%%%%%%%%%%%%%%%%%%%%%%%%%%%%%%%%%%%%%%%%%%%%%%%%%%%%%%%%%%%%%%%%%%%%%%%%%%%%%%%%%%%%%%%%%%%%%%%%%%%%%%%%%%%%%%%%%%%%%%%%%%%%%%%%%%%%%%%%%%%%%%%%%%%%%%%%%%%%%%%%%%
%%%%%%%%%%%%%%%%%%%%%%%%%%%%%%%%%%%%%%%%%%%%%%%%%%%%%%%%%%%%%%%%%%%%%%%%%%%%%%%%%%%%%%%%%%%%%%%%%%%%%%%%%%%%%%%%%%%%%%%%%%%%%%%%%%%%%%%%%%%%%%%%%%%%%%%%%%%%%%%%%%%%%
%%%%%%%%%%%%%%%%%%%%%%%%%%%%%%%%%%%%%%%%%%%%%%%%%%%%%%%%%%%%%%%%%%%%%%%%%%%%%%%%%%%%%%%%%%%%%%%%%%%%%%%%%%%%%%%%%%%%%%%%%%%%%%%%%%%%%%%%%%%%%%%%%%%%%%%%%%%%%%%%%%%%%

\section{The model}\label{sec:model}

%intro, QH state = MN embedding, 
The D3-D7' model is constructed by embedding a probe D7-brane in the background generated by a stack of $N$ D3-branes in such a way that the intersection 
is (2+1)-dimensional \cite{Rey:2008zz} (see also \cite{d3d7flatspace}). As described in \cite{Bergman:2010gm}, this system is a model for the fractional quantum Hall effect.  
For a specific ratio of the charge density to magnetic field and low enough temperature, the D7-brane smoothly ends outside the horizon at some $r_0$. 
This Minkowski embedding holographically corresponds to a quantum Hall state.  

%gap
Minkowski embeddings ordinarily have a gap for charged fluctuations. In the bulk, charges are sourced by open strings stretching from the horizon to 
the tip of the brane; the charge gap is given by the masses of these strings, which is proportional to $r_0$. In \cite{Jokela:2010nu} it was shown that this embedding also has a gap for neutral excitations.

%SL(2,Z) and anyons
However, we showed in \cite{Jokela:2013hta} that the D3-D7' model can be generalized by considering alternative quantizations of the D7-brane gauge field, and for one particular choice, the neutral gap 
closes, giving a superfluid.  The change of the gauge field boundary conditions is implemented by an $SL(2,{\mathbb{Z}})$ electromagnetic  transformation.  
On the boundary, this $SL(2,{\mathbb{Z}})$ action maps from one CFT to another, mixing the charged current and the magnetic field and changing the statistics of the particles.

%electric field -> current
Here, we aim to study the flowing anyon superfluid.  We will start with a quantum Hall state in a background electric field.  Under the  $SL(2,{\mathbb{Z}})$ transformation, 
this electric field will map to a  current.

\subsection{Action and equations of motion}
\label{sec:action}
%background
The metric of the thermal D3-brane background reads:
\be
 L^{-2} ds_{10}^2 = r^2\left(-h(r)dt^2+dx^2+dy^2+dz^2\right)+\frac{1}{r^2}\left(\frac{dr^2}{h(r)}+r^2d\Omega_5^2\right) \ ,
\ee
where the blackening factor $h=1-\left(\frac{r_T}{r}\right)^4$ corresponds to a temperature $T = r_T/(\pi L)$ and the radius of curvature is related to 
the 't Hooft coupling: $L^2=\sqrt{4\pi g_s N}\alpha' = \sqrt{\lambda}\alpha'$. 
We write the metric on the internal sphere as:
\be
 d\Omega_5^2 = d\psi^2+\cos^2\psi\left(d\theta^2+\sin^2\theta\ d\phi^2\right)+\sin^2\psi\left(d\alpha^2+\sin^2\alpha \ d\beta^2\right) \ ,
\ee
with the ranges for angles: $\psi\in [0,\pi/2]$; $\theta,\alpha\in [0,\pi]$; and $\phi,\beta\in [0,2\pi]$. The Ramond-Ramond four-form potential is $C^{(4)}_{txyz} = -L^4 r^4$.

%D7 embedding,stability, internal flux

We embed a flavor D7-brane probe such that it spans $t,x,y$, and $r$, wraps both of the $S^2$'s, and will therefore have a profile $\psi=\psi(r)$ and $z=z(r)$.  
Such an embedding is inherently non-supersymmetric. Indeed, in the decoupling limit,
this is signaled by the presence of a tachyon in the open string spectrum. However, the mass of the tachyon can be lifted above the Breitenlohner-Freedman bound by
turning on a sufficiently large magnetic flux on the internal manifold of the D7-brane worldvolume \cite{Myers:2008me,Bergman:2010gm}, thus rendering the model perturbatively stable.
The same mechanism has also been applied in the T-dual system to stabilize a probe D8-brane in the D2-background \cite{Jokela:2011eb}, where only an infinitesimally 
small internal flux is required.\footnote{In modern language, this configuration can be thought of as D6-branes blown up due to the Myers effect \cite{Kristjansen:2012ny}.}  
We therefore introduce worldvolume flux on one of the internal spheres:
\be
F_{\alpha\beta} = \frac{L^2}{ 2\pi\alpha' } \frac{f}{2}\sin\alpha \ .
\ee
We are concerned here only with Minkowski embeddings, and so we will not turn any flux on the other two-sphere. This also implies the embedding function $z$ will be constant.

%gauge fields for charge, magnetic field, and electric field, physical vs rescaled quantities
In order to construct a quantum Hall state with a background magnetic field, electric field, and charge density, we turn on the following additional components of the worldvolume gauge field:
\bea
F_{xy} &=& B \ =  \ \frac{L^2}{2\pi\alpha'} b\\
F_{tx} &=& E_x  \ =\  \frac{L^2}{2\pi\alpha'} e \\
F_{rt} &=&  \frac{L^2}{2\pi\alpha'} a_t' \ ,
\eea
where the prime represents derivation with respect to $r$.  Furthermore, the electric field will generate a current, so we also include
\bea
F_{rx} &=& \frac{L^2}{2\pi\alpha'} a_x'\\
F_{ry} &=& \frac{L^2}{2\pi\alpha'} a_y'\ .
\eea
We expect that there will be a Hall current in the $y$-direction dual to $a_y'$. But, because the quantum Hall state has vanishing longitudinal conductivity \cite{Bergman:2010gm}, 
there will not be a current in the $x$-direction and we will indeed find
\be
  a_x'=0\  .  
 \ee

%action
In these coordinates, the action of the D7-brane, which consists of a Dirac-Born-Infeld term and a Chern-Simons term, reads \cite{Jokela:2013hta}:
\be
\label{action}
 S = -{\cal N}\int dr \left( r^2 \cos^2\psi \sqrt{f^2+4\sin^4{\psi}} \sqrt Y - c(r)\left(b a'_t + e a_y'\right)  \right) \ ,
\ee
where ${\mathcal N} = 8\pi^2 T_7 V_3 L^5$, $V_3$ is the volume of spacetime, and 
\bea
Y & =& \left(1+ \frac{b^2}{r^4} - \frac{e^2}{hr^4}\right)\left(1+hr^2\psi'^2\right) \nonumber\\
&&- \left(1+ \frac{b^2}{r^4}\right)a_t'^2 + \left(1 - \frac{e^2}{hr^4}\right)h a_y'^2 - \frac{2eb}{r^4}a_t'a_y'  \ .
\eea
The function $c(r)$, essentially representing the axion, is the pullback of the RR four-form potential onto the worldvolume and is given by
\be
 c(r) = \psi+\frac{1}{4}\sin 4\psi-\psi_\infty+\frac{1}{4}\sin 4\psi_\infty \ ,
\ee
where the asymptotic embedding angle $\psi_\infty = \lim_{r\to\infty} \psi(r)$ is related to the internal flux:
\be
 f^2 = 4\sin^2\psi_\infty-8\sin^4\psi_\infty \ .
\ee
In addition a boundary term at the tip of the D7 brane has to be added. This can be seen from either requiring gauge invariance under shifts of the RR four-form potential \cite{Bergman:2010gm} or by consistency of the variation principle.  In our case, this boundary term takes the form
\begin{equation}
S_{boundary}=-{\cal N}c(r_{0})(ba_{t}(r_{0})+ea_{y}(r_{0}))\ ,
\end{equation}
where $r_{0}$ is the smallest $r$ of the D7-brane embedding.
%integrate out gauge fields
In the action (\ref{action}), $a_t$ and $a_y$ are cyclic variables; the associated conserved quantities are the charge density $d \equiv j_t$ %, the longitudinal current $j_x$, 
and Hall current $j_y$.\footnote{The physical charge density and currents, defined by the variation of the on-shell action with respect to the boundary values of $A_t$ and $A_i$, 
are $D= J_t = \frac{2\pi\alpha' \mathcal N}{V_3} d$ and $J_i = \frac{2\pi\alpha' \mathcal N}{V_3} j_i$.} The radial displacement field is $\tilde d(r) = d-2c(r)b$.  
While $d$ gives the total charge density on the boundary, $\tilde d(r)$ measures how much of that charge is due to sources in the bulk located at radial positions below $r$. 
Similarly, $j_y$ is the total Hall current, and $\tilde j_y(r) =  j_y-2c(r) e$ is the current due sources below $r$.

%Cartesian coordinates
With an eye toward the fluctuation analysis, we will consider Cartesian-like coordinates $(R, \rho)$ instead of the polar coordinates $(r, \psi)$:
\bea
 \rho & = & r\sin\psi \\
 R    & = & r\cos\psi \ .
\eea
The embedding is now described by $\rho=\rho(R)$, where $R$ is the new worldvolume coordinate. We will still write $r$ explicitly in the equations to follow, 
but it should  be read as $r=\sqrt{\rho(R)^2+R^2}$.  Until now prime has denoted derivative with respect to $r$; from now on, it will instead indicate a derivative with respect to $R$.

Performing an appropriate Legendre transformation to eliminate the cyclic variables, we obtain the following action (including the appropriately mapped boundary term) for the embedding field $\rho(R)$:
\be\label{Ruthian}
 S = -{\cal N} \int\frac{dR}{h r}\sqrt{(\rho\rho'+R)̂^2+h(R\rho'-\rho)^2}\sqrt{X} \ ,
\ee
where 
\bea
   X & = & h\left(1+\frac{b^2}{r^4}-\frac{e^2}{hr^4}\right)\left(4hR^4\left(f^2+4\frac{\rho^4}{r^4}\right)+h\tilde d(R)^2-\tilde j_y(R)^2\right) \nonumber\\
   && -\frac{(hb\tilde d(R)- e\tilde j_y(R))^2}{r^4} \ .\label{eq:X}
\eea
The solutions for the gauge fields are:
\bea
\label{ateom}
 a'_t & = & \left(h\tilde d\left(1-\frac{e^2}{hr^4}\right)+\frac{eb}{r^4}\tilde j_y\right)\sqrt{\frac{(\rho\rho'+R)^2+h(R\rho'-\rho)^2}{r^2 X}} \\
 %\label{axeom}
 %a'_x & = & -j_x\left(1+\frac{b^2}{r^4}-\frac{e^2}{hr^4}\right)\sqrt{\frac{(\rho\rho'+R)^2+h(R\rho'-\rho)^2}{r^2 X}} \\
\label{ayeom}
 a'_y & = & \left(\frac{e \tilde d b}{r^4}-\left(1+\frac{b^2}{r^4}\right)\tilde j_y\right)\sqrt{\frac{(\rho\rho'+R)^2+h(R\rho'-\rho)^2}{r^2 X}}  \ .
\eea
We obtain a complete solution by first numerically solving the equation of motion for $\rho(R)$ derived from (\ref{Ruthian})  and then using this to numerically integrate (\ref{ateom}) and (\ref{ayeom}).

%mass and Delta_\pm
The mass\footnote{Here, $m$ is related to the physical mass by $M = 2\pi\alpha' m$.} of the fermions is extracted from the leading UV behavior of the embedding:
\be
m = r^{-\Delta_+}\sin\left(\arctan\left(\frac{\rho}{R}\right)-\psi_\infty \right) \ ,
\ee
where the corresponding operator dimensions are
\be
 \Delta_\pm = -\frac{3}{2}\pm\frac{1}{2}\sqrt{73-\frac{48}{\cos^2\psi_\infty}} \ .
\ee
In this paper, we fix $f=\frac{2}{3}$, which yields $\psi_\infty = \frac{1}{2} \arccos\left(\frac{1}{3}\right)$.  This choice leads to zero anomalous mass dimension for the fermions; 
this is, $\Delta_+ = -1$. We do not expect to find qualitatively different results for different values of $f$.

\subsection{Minkowski embeddings}
\label{sec:MNembeddings}

%definition
Probe brane embeddings can be classified into two categories.  Generically, probe branes cross the horizon; these are black hole embeddings. They are interesting in myriad ways  \cite{Davis:2011gi,Bergman:2011rf,Jokela:2012vn}. 
However, in special cases, 
probe branes can end smoothly at some $r_0$ above the horizon as one of the wrapped $S^2$'s shrinks to zero size, yielding Minkowski (MN) embeddings, which are what we focus on here.

%fixed d/b, Hall current
There are constraints coming from the requirement that the embeddings are of MN type. Essentially, we are demanding that there are no sources at the tip of the D7-brane. 
Due to the effects of the Chern-Simons term, this does not mean that we need
to require the charge density to vanish. Rather, via a mechanism revealed in \cite{Bergman:2010gm}, we require that the charge density be locked with the magnetic field in such a way that the
radial displacement field is forced to vanish at the tip:  $\tilde d(R=0) = 0$. A similar argument holds for the  currents: $j_x = \tilde j_y(R=0) =0$. These conditions yield:
\bea
\label{lockingcondition}
  \frac{d}{b} & = & 2c(R=0) = \pi -2\psi_\infty+\frac{1}{2}\sin 4\psi_\infty\equiv \frac{\pi\nu}{N} \\ 
    j_{y}    & = & \frac{d}{b}e \ ,
\eea
of which the former dictates the filling fraction $\nu$, which ultimately follows from the amount of flux $f$ we turned on.  As expected, the only nonzero current is the Hall current, 
and the conductivities are precisely those of a quantum Hall state \cite{Bergman:2010gm}:
\bea
\sigma_{xx} &=& 0 \\
\sigma_{xy} & =& \frac{\nu}{2\pi} \ .
\eea

\subsection{Rescaled variables}
%scaling out b
We have the freedom to scale out some parameters. Since the system is pretty robust against temperature variations, it is better to scale out the magnetic field as follows:
\be
 R = \sqrt b \tilde R \ , \ \rho = \sqrt b \tilde\rho \ , \ r = \sqrt b\tilde r \ , \ a_\mu = \sqrt b\tilde a_\mu \ .
\ee
In terms of the rescaled variables, the action (\ref{Ruthian}) has the same form, but with tildes added to all quantities and with the overall normalization:
\be
\tilde {\cal N} = {\cal N} b^{3/2} \ .
\ee

%parameter space
We are then left with four independent parameters; these are the reduced temperature, the electric field, and the mass of the fermions:
\bea
 \tilde r_T & = & \frac{r_T}{\sqrt b} \\
 \tilde e & = & \frac{e}{b} \\
 \tilde m & =& b^{\Delta_+/2} m \ ,
\eea
along with the internal flux $f$.  As was mentioned in Sec.~\ref{sec:action}, different values of $f$ yield qualitatively similar results, and we will therefore fix $f=\frac{2}{3}$.
Moreover, it turns out that different $\tilde m$ do not induce qualitative changes either; we therefore fix $\tilde m=-8$ in what follows. In practice, 
we thus have a two-dimensional parameter space  $(\tilde r_T,\tilde e)$ to explore.

%%%%%%%%%%%%%%%%%%%%%%%%%%%%%%%%%%%%%%%%%%%%%%%%%%%%%%%%%%%%%%%%%%%%%%%%%%%%%%%%%%%%%%%%%%%%%%%%%%%%%%%%%%%%%%%%%%%%%%%%%%%%%%%%%%%%%%%%%%%%%%%%%%%%%%%%%%%%%%%%%%%%%
%%%%%%%%%%%%%%%%%%%%%%%%%%%%%%%%%%%%%%%%%%%%%%%%%%%%%%%%%%%%%%%%%%%%%%%%%%%%%%%%%%%%%%%%%%%%%%%%%%%%%%%%%%%%%%%%%%%%%%%%%%%%%%%%%%%%%%%%%%%%%%%%%%%%%%%%%%%%%%%%%%%%%
%%%%%%%%%%%%%%%%%%%%%%%%%%%%%%%%%%%%%%%%%%%%%%%%%%%%%%%%%%%%%%%%%%%%%%%%%%%%%%%%%%%%%%%%%%%%%%%%%%%%%%%%%%%%%%%%%%%%%%%%%%%%%%%%%%%%%%%%%%%%%%%%%%%%%%%%%%%%%%%%%%%%%
%%%%%%%%%%%%%%%%%%%%%%%%%%%%%%%%%%%%%%%%%%%%%%%%%%%%%%%%%%%%%%%%%%%%%%%%%%%%%%%%%%%%%%%%%%%%%%%%%%%%%%%%%%%%%%%%%%%%%%%%%%%%%%%%%%%%%%%%%%%%%%%%%%%%%%%%%%%%%%%%%%%%%
%%%%%%%%%%%%%%%%%%%%%%%%%%%%%%%%%%%%%%%%%%%%%%%%%%%%%%%%%%%%%%%%%%%%%%%%%%%%%%%%%%%%%%%%%%%%%%%%%%%%%%%%%%%%%%%%%%%%%%%%%%%%%%%%%%%%%%%%%%%%%%%%%%%%%%%%%%%%%%%%%%%%%

\section{Flowing superfluid}
\label{sec:flowingsuperfluid}

%SL(2,Z) from QH
\subsection{$SL(2,{\mathbb{Z}})$ transformations}\label{sec:SL2Z}

%CFT
For any CFT in 2+1 dimensions with a conserved $U(1)$ charge, there are two natural operations which can transform it into another CFT:  adding a Chern-Simons term for an external vector field and making an external vector field dynamical.  Together, these operations generate an $SL(2,{\mathbb{Z}})$ action transforming one CFT into another \cite{Witten:2003ya, Burgess and Dolan}. Holographically, this mapping corresponds to changing the boundary conditions of the bulk gauge field and thereby imposing alternative quantization.

%natural SL(2,Z) action on CFTs, review SL(2,Z) from last paper, alternative quantization from added boundary terms, transformation of J and B
In \cite{Jokela:2013hta}, we described the D3-D7' model with alternative quantization of the bulk gauge field.  
Standard quantization corresponds to Dirichlet boundary conditions for the gauge field $A^\mu$.  The variation of the renormalized on-shell action is just a boundary term:
\be
\delta S_{D}=\int _{boundary} J_{\mu}\delta A^{\mu} \ ,
\ee
where $J_{\mu}=\frac{\delta S_{D}}{\delta A^{\mu}}$ is the conserved $U(1)$ current in the CFT.  This implies that $A^\mu$ must be kept fixed at the boundary; that is, $\delta A^\mu = 0$.  
To implement mixed boundary conditions, we can add a general boundary term to the action.   Defining, up to gauge transformations, a vector $V^\mu$ such that
\be
J_{\mu}=\frac{1}{2\pi}\epsilon_{\mu \rho \nu}\partial^{\rho}V^{\nu} \ ,
\ee
the most general boundary term we can add takes the form
\be
S_{gen}=S_{D}+\frac{1}{2\pi}\int_{boundary}[a_1\epsilon_{\mu \rho \nu}A^{\mu}\partial^{\rho}V^{\nu}+a_2\epsilon_{\mu \rho \nu}A^{\mu}\partial^{\rho}A^{\nu}+a_3\epsilon_{\mu \rho \nu}V^{\mu}\partial^{\rho}V^{\nu}]
\ee
for arbitrary $a_1$, $a_2$, and $a_3$.   The variation of the on-shell action can be written as
\be
\label{deltaSgen}
\delta S_{gen}=\int_{boundary} (a_{s}J_{\mu}+b_{s} {\cal B}_{\mu})(c_{s}\delta V^{\mu}+d_{s} \delta A^{\mu}) \ ,
\ee
where 
\be
{\cal B}_{\mu}=\frac{1}{2\pi}\epsilon_{\mu \rho \nu}\partial^{\rho}A^{\nu} \ ,
\ee
and where
\be
\label{sl2zrleations}
a_sd_s=1+a_1, \ \ b_sc_s=a_1,\ \ b_s d_s=2a_2, \ \ a_s c_s=2a_3 \ .
\ee
Note that (\ref{sl2zrleations}) implies that $a_{s}d_{s}-b_{s}c_{s}=1$.  The parametrization of (\ref{deltaSgen}) makes clear that  
\be
\left(\begin{array}{cc}a_s & b_s \\c_s & d_s\end{array}\right) \in SL(2,{\mathbb{R}}) \ ,
\ee
and we recognize this change of boundary conditions as an $SL(2,{\mathbb{R}})$ transformation mapping the original boundary theory into a new one.     
The new boundary condition fixes $\mathcal B^*_\mu$, and the new conserved current is $J^*_\mu$.  These are related to the original variables by an $SL(2,{\mathbb{R}})$ transformation:
\be
\label{SL2Zofcurrents}
\left(\begin{array}{c}J^*_\mu \\  {\cal B}^*_\mu \end{array}\right) =
\left(\begin{array}{cc}a_s & b_s \\c_s & d_s\end{array}\right) \left(\begin{array}{c}J_\mu \\  {\cal B}_\mu \end{array}\right) \ . 
\ee
Because charges in the bulk theory are quantized, we are, in fact, restricted to transformations in the subgroup $SL(2,{\mathbb{Z}})$.

%How to get superfluid, SL(2,Z) of QH
An anyonic superfluid state can be obtained by a judicious $SL(2,{\mathbb{Z}})$ transformation from a quantum Hall state in standard quantization.  
A quantum Hall state with filling fraction $\nu = \frac{2\pi D}{B}$ and background electric field $E_x$ has a Hall current $J_y = \frac{D}{B} E_x$.   
A general $SL(2,{\mathbb{Z}})$ transformation of such a quantum Hall state gives, via (\ref{SL2Zofcurrents}),
\begin{eqnarray}
{J}^{*}_{x}&=0 \hspace{2.3cm} &{E}^{*}_{y}=0 \nonumber\\
{J}^{*}_{y}&=a_s J_{y} -b_s \frac{E_{x}}{2\pi} \hspace{0.5cm} &-\frac{{E}^{*}_{x}}{2\pi}=c_s J_{y} -d_s \frac{E_{x}}{2\pi} \nonumber\\
-{D}^{*}&=-a_s D +b_s \frac{B}{2\pi} \hspace{0.5cm} &\frac{{B}^{*}}{2\pi}=-c_s D +d_s \frac{B}{2\pi} \ .
\label{jetrans}
\end{eqnarray}

%SL(2,Z) for superfluid, why this is a superfluid
To end up with a superfluid, we need a final state with no electric or magnetic field.  We therefore choose a transformation with 
\be
\frac{d_{s}}{c_s}=\nu = \frac{2\pi D}{B} \ ,
\ee
which implies $B^{*}=0$.  Since we started with a quantum Hall state with $J_y = \frac{D}{B} E_x$,  this choice implies $E^{*}_x=0$, as well.  The current and charge density in the new state are
\bea
J^{*}_{y} & = & \left(a_s \frac{D}{B}-\frac{b_s}{2\pi} \right)E_{x} \\
 D^{*} & = & \left(a_s \frac{D}{B}-\frac{b_s}{2\pi}\right)B \ .
\eea
%The velocity of the current is then
%\be
%v_{f}=\frac{J^{*}_{y}}{D^{*}}=\frac{E_{x}}{B}=\tilde{e} \ .
%\ee
The new state has a persistent current in the absence of a driving electric field, as would be expected for a superfluid.  In \cite{Jokela:2013hta}, we also showed that, in the case with $\tilde e = 0$, this $SL(2,{\mathbb{Z}})$ mapping produced a superfluid with the requisite gapless mode.  We will further investigate the dispersion of this mode in Sec.~\ref{sec:fluctuations}. 

At nonzero temperature, a superfluid can be described as a mixture of two components, the superfluid and an ordinary fluid of thermally excited phonons.   Both components contribute to the current $J$, and the velocity $v_{ave} = J/D$ gives a weighted average of the superfluid and normal fluid velocities.

At zero temperature, the normal fluid is absent.  In this case  the superfluid velocity is
\be
\label{vfzerotemp}
v_f=\frac{J^{*}_{y}}{D^{*}}=\frac{E_{x}}{B}=\tilde{e} \ .
\ee

%superfluid velocity, holographic conventional superfluid velocity, v_f dual to Jy, boundary value of Ay, still the same for anyonic superfluid, v_f = ..., T=0 limit
The velocity of a conventional superfluid is given by the gradient of the order parameter.  In holographic superfluids \cite{Basu:2008st, Herzog:2008he}, 
the superfluid velocity is the dual source for the current $J^i$ and is therefore given by the boundary value of the bulk gauge field $A_i$.  For a superfluid flowing in the $y$-direction, the velocity is
\be
\label{vfdefinition}
v_f = \left. - \frac{A_y}{A_{t}} \right|_{boundary} \ .
\ee
Anyonic superfluids are not characterized by a local order parameter.  However, it seems natural that the superfluid velocity is still given by (\ref{vfdefinition}) with $A_{\mu} \rightarrow A_{\mu}^{*}$, where
$A_{\mu}^{*}=c_s V_{\mu}+d_sA_{\mu}$ is the $SL(2,{\mathbb{Z}})$ transformed gauge field. In our case we have a nonzero $J_{y}$ and a nonzero $J_{t}$, so we can pick a gauge where  $V_{y}=V_{t}=0$. We can also choose a gauge where the electric and magnetic field come from $A_{x}$. In this case we can then compute
\be
\label{vfdefinition1}
v_f = \left.- \frac{A_{y}^{*}}{A_{t}^{*}} \right|_{boundary} =\left. -\frac{a_{y}}{a_{t}} \right|_{boundary}\ .
\ee

The boundary values of $a_t$ and $a_y$ can be found by integrating the expressions (\ref{ateom}) and (\ref{ayeom}) for $a_t'$ and $a_y'$:
\bea
\label{atintegral}
a_t(\infty) &=& \int_0^\infty dR \  h \tilde d \sqrt{\frac{(\rho\rho'+R)^2+h(R\rho'-\rho)^2}{r^2 X}} + a_t(0) \\
\label{ayintegral}
a_y(\infty) &=& - \frac{e}{b}  \int_0^\infty dR \  \tilde d \sqrt{\frac{(\rho\rho'+R)^2+h(R\rho'-\rho)^2}{r^2 X}} + a_y(0)\ ,
\eea
which, for a given embedding $\rho(R)$, can be integrated numerically.  One subtlety with expressions (\ref{atintegral}) and (\ref{ayintegral}) is the IR boundary condition at $R=0$.  
For a MN embedding, the value of the gauge field at the tip is not fixed.  For example, as was seen in various contexts \cite{Ghoroku:2007re, Mateos:2007vc, Bergman:2010gm, Jokela:2011eb}, the charge density in a MN embedding is completely independent of the chemical potential.  
In order to obtain the correct zero-temperature limit (\ref{vfzerotemp}), we must choose 
\be
 a_\mu(R=0) = 0 \ ,
\ee
which fixes the IR boundary terms in (\ref{atintegral}) and (\ref{ayintegral}).

Thus the expression (\ref{vfdefinition1}), using our chosen boundary conditions,  reduces at zero temperature to $v_f = \frac{e}{b} = \tilde e$, as desired.  More generally, from (\ref{atintegral}) and (\ref{ayintegral}) and the fact that $h \leq 1$, we find that $v_f \geq \tilde e$.  We can thus parametrize our flowing solutions by $v_{f}$ or by $\tilde{e}$.

\subsection{Superflowing solutions}\label{sec:backgroundnumerics}

%solve for MN embeddings
To find solutions corresponding to flowing superfluids, we numerically solve the equations for the D7-brane embedding and gauge fields in the 
presence of worldvolume electric and magnetic fields.  We focus here only on MN embeddings, so we need to impose the IR boundary conditions discussed in Sec.~\ref{sec:MNembeddings}.

%bulk solutions independent of choice of quantization
Note that, because different quantizations only differ by boundary terms, the bulk equations of motion are independent of the choice of quantization.  
After an $SL(2,{\mathbb{Z}})$ transformation, the solutions of the equations of motion are therefore the same, but their physical interpretation is different.  
In standard quantization, $\tilde{e}$ is a background electric field, while in the alternative quantization appropriate to the anyonic superfluid, it is the average fluid velocity $v_{ave}$.

\begin{figure}[!ht]
\center
\includegraphics[width=0.7\textwidth]{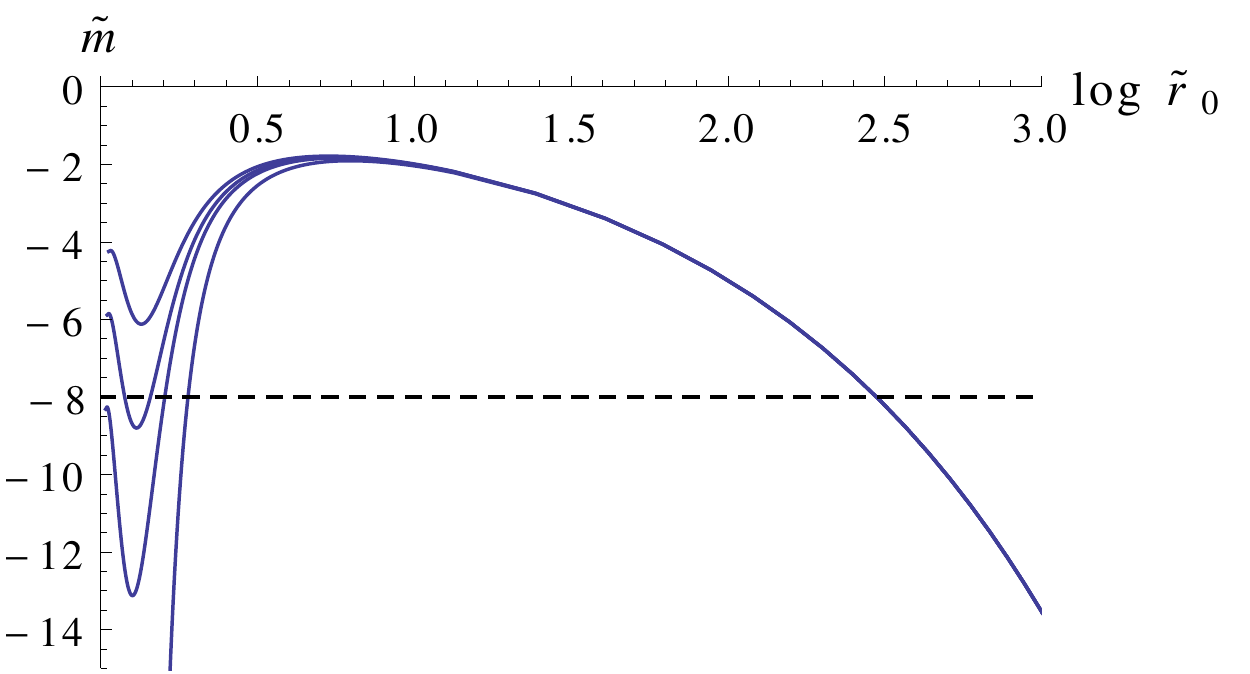}
\caption{The different mass curves for MN solutions showing the relationship between $\log \tilde r_0$ and $\tilde m$, at $\tilde r_T=1$ and (from bottom) $\tilde e= 0,0.35,0.4,0.45,$ and 0.6. 
The dashed black line at $\tilde m=-8$ is to guide the eye; this is the mass we will fix in the following sections. The thermodynamically preferred solution is 
on the branch with positive slope.  %In addition to the solutions shown here, there is another unstable  $\tilde m=-8$ solution at large $\tilde r_0$ \cite{Jokela:2010nu} not visible in this plot. 
 Notice that for $\tilde e \gtrsim 0.41$, there are no stable $\tilde m = -8$ solutions.} 
\label{fig:embeddings} 
\end{figure}

\begin{figure}[!ht]
\center
\includegraphics[width=0.7\textwidth]{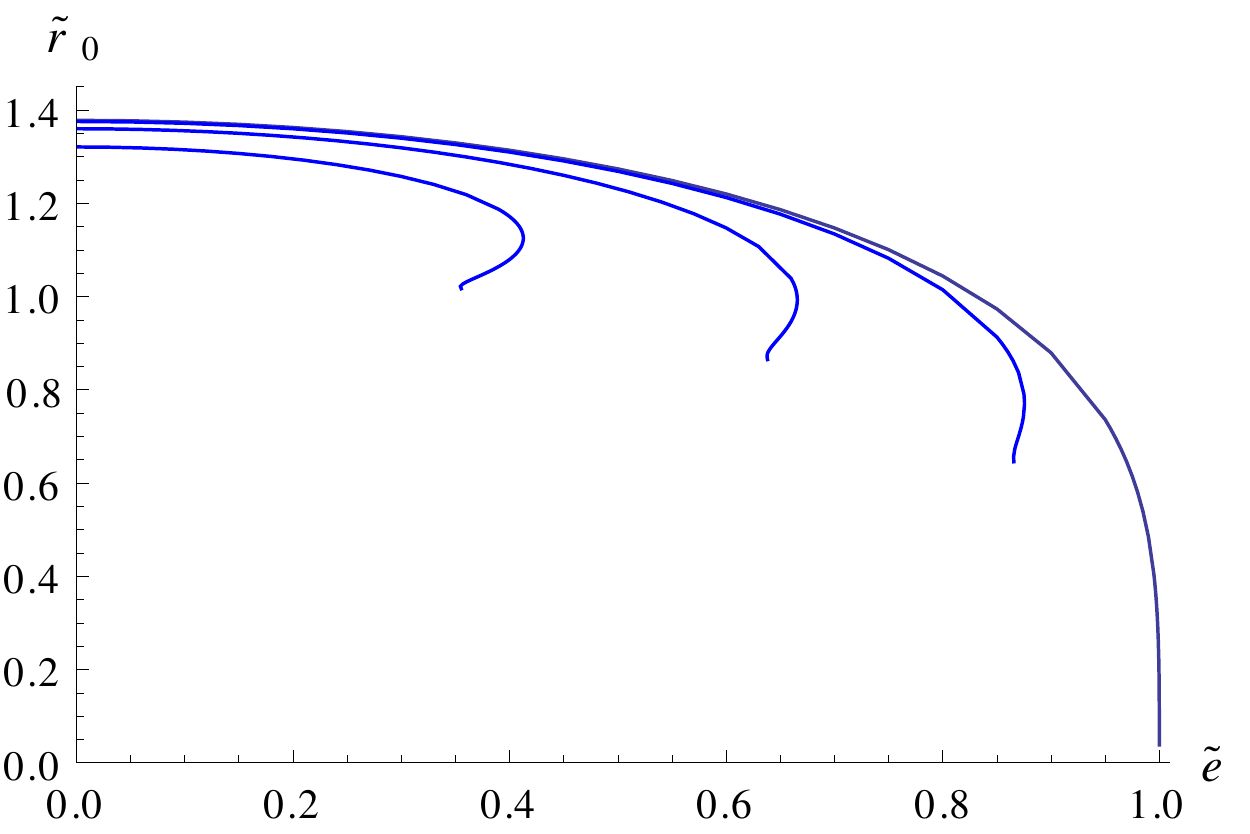}
\caption{The mass gap $\tilde r_0=\tilde\rho_0$ as a function of $\tilde e$ at fixed $\tilde m=-8$, for several temperatures: 
From right to left, the curves correspond to $\tilde r_T=0,0.5,0.8,$ and 1.
At $\tilde r_T=0$, the gap smoothly closes at $\tilde e=1$. As the temperature is increased, the maximum $\tilde e$ 
decreases: $\tilde e_{max} = 0.91, 0.67, 0.41$ for $\tilde r_T = 0.5,0.8,1$, respectively.  
Furthermore, for $\tilde r_T > 0$, additional solutions appear at $\tilde e \lesssim \tilde e_{max}$, which presumably are not thermodynamically preferred. }\label{fig:gap} 
\end{figure}

%results: multiple r_0 for fixed m, pick stable one (fig 1), for stable solutions r_0 decreases with e (can be seen in Fig 1, but plotted in fig 2)
For fixed values of $\tilde m$ and $\tilde e$, there are, in general, multiple MN solutions with different values of $\tilde r_0$, as shown in Fig.~\ref{fig:embeddings}.  
It was found in \cite{Bergman:2010gm} that for $\tilde e=0$, the thermodynamically preferred solution\footnote{For the $SL(2,{\mathbb{Z}})$ transformed case, 
there are boundary terms that need to be taken into account, but for all MN embeddings, these extra terms have the same value.} 
was the one with second-largest $\tilde r_0$; in \cite{Jokela:2010nu} it was also shown to be perturbatively stable. We believe that this thermodynamic argument holds at $\tilde e>0$. %and $\frac{d\tilde m}{d\tilde r_0} > 0$. 
From Fig.~\ref{fig:embeddings}, it is evident that for a given choice of $\tilde m$, when $\tilde e$ becomes sufficiently large, there are no solutions (apart from an unstable branch of solutions 
for which $\tilde r_0\gtrsim 2$).  
This can be seen even more clearly in Fig.~\ref{fig:gap} which shows $\tilde r_0$ as a function of $\tilde e$ for a fixed $\tilde m$.  
For each temperature, there is a maximum $\tilde e$ such that there are no relevant MN solutions for $\tilde e > \tilde e_{max}$.  
In the standard quantization, $\tilde e_{max}$ is the maximum electric field.  For the anyonic superfluid, the physical interpretation is that there is a maximum velocity $v_{max}$ beyond which no relevant solutions exist.

\begin{figure}[!ht]
\center
\includegraphics[width=0.45\textwidth]{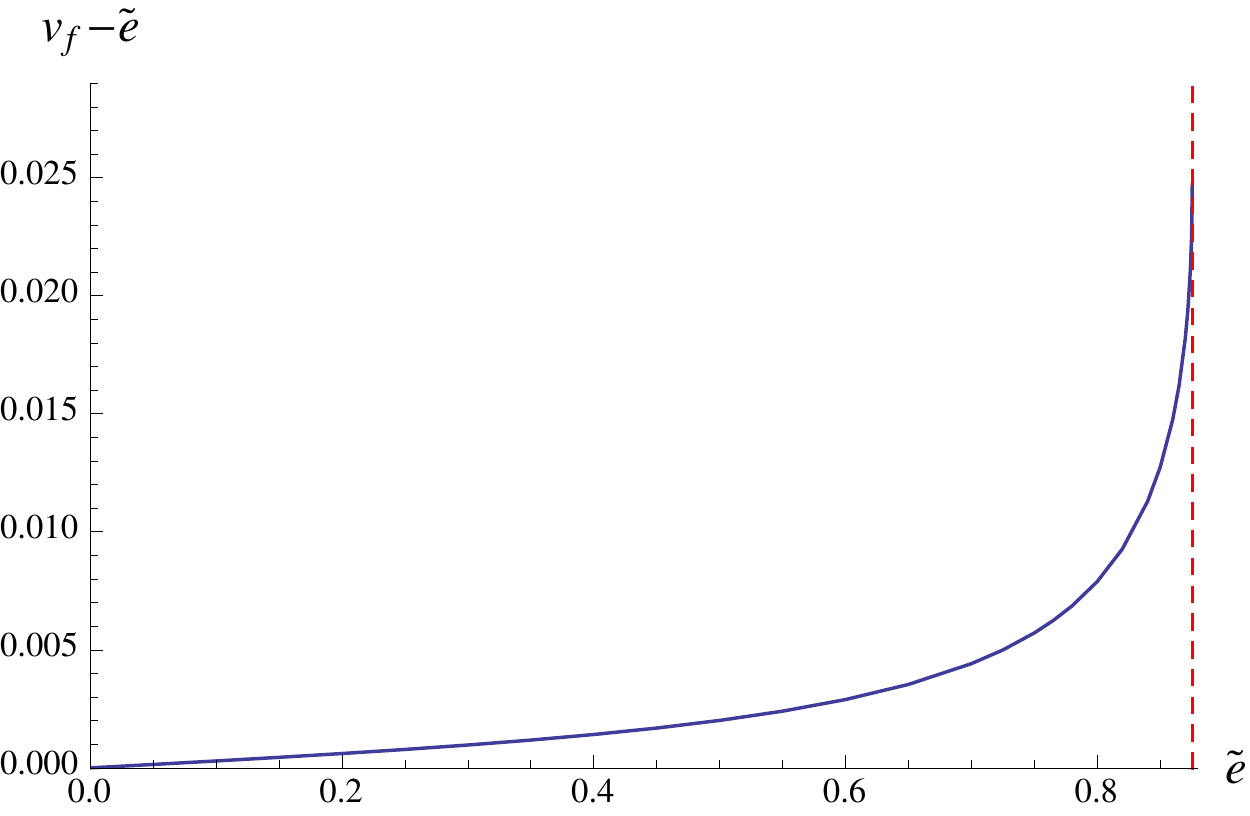}
\includegraphics[width=0.45\textwidth]{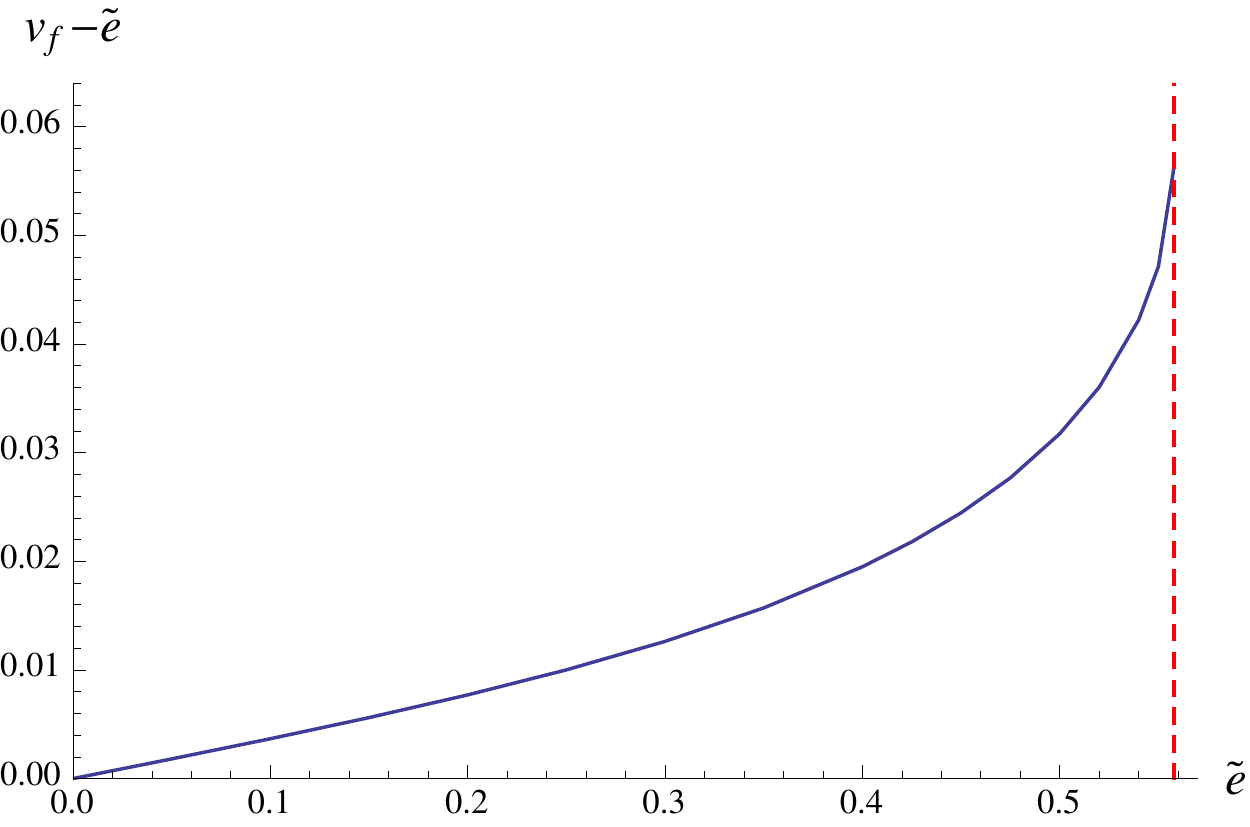}
\caption{The difference between $v_f$ and $\tilde e$ plotted as a function of $\tilde e$ at $\tilde m=-8$ and at (left) $\tilde r_T = 0.5$ and (right) $\tilde r_T=0.9$.  The curves terminate at (left) $\tilde e_{max} = 0.876$ and (right) $\tilde e_{max} = 0.558$, which are indicated by the dashed red lines. Note that, as expected, $v_f \geq \tilde e$.}\label{fig:vf-e_VS_e} 
\end{figure}

%v_f vs e, Fig v-e VS e
For a given MN solution, the superfluid velocity can be computed numerically via (\ref{vfdefinition1}).  We find that, in general, $v_f$ is slightly larger than $\tilde e$.  We plot the difference $v_f - \tilde e$ for two fixed temperatures in Fig.~\ref{fig:vf-e_VS_e}.  The difference is largest as $\tilde e$ approaches $\tilde e_{max}$.

%e_max vs T
At zero temperature $\tilde e_{max} = v_{max} = 1$, which simply means that the superfluid can not flow faster than the speed of light.  
In this case, as the superfluid velocity increases to $v_{max}$, the charge mass gap decreases, {\emph{i.e.}} $\tilde{r}_{0} \rightarrow 0$ for all $\tilde{m}$,   
as $\tilde e = v_f \to 1$; see Fig.~\ref{fig:gap}.  At nonzero temperatures, both $\tilde e_{max} < 1$ and $v_{max} < 1$, and both depend on $\tilde{m}$.  In addition, at nonzero temperature the charge 
mass gap does not close as $\tilde e  \to \tilde e_{max}$.  Instead, another branch of solutions with smaller $\tilde r_0$ appears and merges with the thermodynamically preferred stable solution at $\tilde e = \tilde e_{max}$.

We will see by analyzing the fluctuation spectrum, however, that the anyonic superfluid becomes unstable before the maximum superfluid velocity is reached.

%%%%%%%%%%%%%%%%%%%%%%%%%%%%%%%%%%%%%%%%%%%%%%%%%%%%%%%%%%%%%%%%%%%%%%%%%%%%%%%%%%%%%%%%%%%%%%%%%%%%%%%%%%%%%%%%%%%%%%%%%%%%%%%%%%%%%%%%%%%%%%%%%%%%%%%%%%%%%%%%%%%%%
%%%%%%%%%%%%%%%%%%%%%%%%%%%%%%%%%%%%%%%%%%%%%%%%%%%%%%%%%%%%%%%%%%%%%%%%%%%%%%%%%%%%%%%%%%%%%%%%%%%%%%%%%%%%%%%%%%%%%%%%%%%%%%%%%%%%%%%%%%%%%%%%%%%%%%%%%%%%%%%%%%%%%
%%%%%%%%%%%%%%%%%%%%%%%%%%%%%%%%%%%%%%%%%%%%%%%%%%%%%%%%%%%%%%%%%%%%%%%%%%%%%%%%%%%%%%%%%%%%%%%%%%%%%%%%%%%%%%%%%%%%%%%%%%%%%%%%%%%%%%%%%%%%%%%%%%%%%%%%%%%%%%%%%%%%%

\section{Fluctuations}\label{sec:fluctuations}

%intro: like previous paper and roton paper, but with nonzero superfluid velocity 
We now compute the spectrum of collective excitations of the anyon superfluid flowing with velocity $v_f$.  
The $SL(2,{\mathbb{Z}})$ transformation used in Sec.~\ref{sec:SL2Z} to generate the superfluid state acts in the bulk to change the 
boundary conditions, and thereby, the quantization of the gauge field.  We must, therefore, analyze the fluctuations of all fields around the MN background, 
imposing alternative boundary conditions on the bulk gauge fields, as previously described in \cite{Jokela:2013hta}.  

\subsection{Set up}
%wavelike ansatz, gauge fixing, gauge invariant variables
In the case of a nonflowing, isotropic superfluid \cite{Jokela:2013hta}, one could use rotational symmetry 
in the $x-y$ plane to align the fluctuation with, say, the $x$-axis.  However, the superfluid flowing in the $y$-direction breaks this symmetry.  
The excitation frequency will in fact depend on the relative angle between the superflow and the fluctuation.  

We impose the following wavelike ansatz on the fluctuations:\footnote{The $\delta \tilde z$ fluctuation is completely decoupled, and we will not focus on it.}
\bea
\delta \tilde a_\mu &=&  \delta \tilde a_\mu(\tilde R) e^{-i\omega t + i k_x x + i k_y y} \\
\delta \tilde \rho &=&  \delta \tilde \rho(\tilde R) e^{-i\omega t + i k_x x + i k_y y} \  .
\eea
The rescaled frequency and momenta are
\be
(\omega,k_x, k_y) = \sqrt{b} (\tilde\omega, \tilde k_x,  \tilde k_y) \ .
\ee
It is preferable, however, to work with the gauge-invariant field strength perturbations:
\bea
\label{exdef}
\delta \tilde e_x &=&  \tilde \omega\delta \tilde a_x + \tilde k_x \delta \tilde a_t \\
\label{eydef}
\delta \tilde e_y &=&  \tilde \omega\delta \tilde a_y + \tilde k_y \delta \tilde a_t \\
\label{bdef}
\delta \tilde b &=& \tilde k_x \delta \tilde a_y - \tilde k_y \delta \tilde a_x \ .
\eea

With this ansatz, we expand the D7-brane action (\ref{Ruthian}) to second order and derive the 
equations of motion for $\delta \tilde e_x$, $\delta \tilde e_y$, $\delta \tilde b$, and $\delta \tilde \rho$.  This system of coupled, linear ordinary differential equations is extremely long and not particularly illuminating;  we will therefore not reproduce it here.

%%%%%%%%%%%%%%%%%%%%%%%%%%%%%%%%%%%%%%%%%%%%%%%%%%%%%%%%%%%%%%%%%%%%%%%%%%%%%%%%%%%%%%%%%%%%%%%%%%%%%%%%%

\subsection{Alternative boundary conditions}

%alternative boundary conditions, general mixed conditions with n
The general mixed boundary conditions for fluctuations of the gauge field can be written as:
\be
\label{alternativeboundarycondition}
0 = -n \, \delta F_{\mu u} + \frac{1}{2} \epsilon_{\mu\nu\rho} \delta F^{\nu\rho} \ ,
\ee
where indices $\mu, \nu$, and $\rho$ are (2+1)-dimensional boundary coordinates, raised and lowered by the 
flat metric $\eta_{\mu\nu}$, and the inverse radial coordinate $u = 1/r$.  The boundary is  therefore located at $u=0$.  
The parameter $n$ indicates the particular choice of quantization (see also \cite{Brattan:2013wya} for more discussion on this).  
The standard quantization with Dirichlet boundary conditions corresponds to $n=0$, and $n=\infty$ gives Neumann boundary conditions.

%relating n to SL(2,Z) parametrization, maybe more detail than needed?
The parameter $n$ is related to the $SL(2,{\mathbb{Z}})$ parametrization in Sec.~\ref{sec:SL2Z}.  From (\ref{SL2Zofcurrents}), we see that the new boundary condition, $\delta B^* = 0$, implies mixed fluctuations in terms of the original charge and magnetic field.  The charge $d$ and magnetic field $b$ are related to the physical charge $D$ and magnetic field $B$ by
%maybe move this earlier?
\be
\frac{D}{B} =  \frac{(2\pi\alpha')^2 \mathcal N}{L V_{2,1}} \frac{d}{b} =  \frac{N}{2\pi^2} \frac{d}{b} \ ,
\ee
and the charge is related to the boundary value of the bulk gauge field via
\be
F_{0 u} (u=0)=\frac{d}{\sqrt{4\cos^4 \psi_{\infty}(f^2+4\sin^4 \psi_{\infty})}} \ .
\ee
Writing (\ref{SL2Zofcurrents}) in terms of the bulk gauge field and comparing the result to (\ref{alternativeboundarycondition}) gives
\be
\label{ndef}
 n = \frac{N}{\pi} \sqrt{4\cos^4 \psi_{\infty}(f^2+4\sin^4 \psi_{\infty})} \  \frac{c_s}{d_s} \ .
\ee

%n for superfluid
As explained in Sec.~\ref{sec:SL2Z}, the superfluid phase is obtained from the quantum Hall phase by an $SL(2,{\mathbb{Z}})$ transformation for which $d_s/c_s = \nu$, where the filling fraction $\nu$ is given in terms of $\psi_\infty$ by (\ref{lockingcondition}).  In the superfluid phase, therefore, the gauge field fluctuations obey boundary conditions with
\be
n =  \frac{\sqrt{4\cos^4 \psi_{\infty}(f^2+4\sin^4 \psi_{\infty})}}{\pi - 2\psi_\infty + \frac{1}{2}\sin(4\psi_\infty)} \ .
\ee

%gauge invariant variables
In order to write these boundary conditions entirely in terms of gauge-invariant quantities, we can use the gauge constraint coming from the equation of motion for $\delta a_u$ which, for $u \rightarrow 0$, reads 
\be
 \omega \partial_u \delta a_0 + k_x \partial_u \delta a_x + k_y \partial_u \delta a_y = 0 \ .
\ee
The boundary condition (\ref{alternativeboundarycondition}) can then be written as follows:
\bea
-n \, \frac{k_x \partial_u \delta e_x + k_y \partial_u \delta e_y}{\omega^2 - k_x^2 - k_y^2} + i \delta b &=& 0 
\label{mu=0} \\
n \, \frac{\left(\omega^2 - k_y^2\right) \partial_u \delta e_x + k_x k_y \partial_u \delta  e_y}{\omega \left(\omega^2 - k_x^2 - k_y^2\right)} - i \delta e_y &=& 0
 \label{mu=1} \\
n \, \frac{k_x k_y \partial_u \delta  e_x + \left(\omega^2 - k_y^2\right)  \partial_u \delta e_y}{\omega \left(\omega^2 - k_x^2 - k_y^2\right)} + i \delta e_x &=& 0 
\label{mu=2} \ .
\eea
Now we need to put the boundary conditions (\ref{mu=0}), (\ref{mu=1}), and (\ref{mu=2}) in terms of the rescaled radial coordinate $\tilde R$  and the other rescaled variables for use in the numerical calculations:
\bea
n \, \frac{\tilde R^2}{\cos \psi_\infty} \frac{\tilde k_x \partial_{\tilde R} \delta \tilde e_x + \tilde k_y \partial_{\tilde R} \delta \tilde e_y}{\tilde \omega^2 - \tilde k_x^2 - \tilde k_y^2} + i \tilde \delta b &=& 0 
\nonumber \\
n \, \frac{\tilde R^2}{\cos \psi_\infty} \frac{\left(\tilde \omega^2 - \tilde k_y^2\right) \partial_{\tilde R} \delta \tilde e_x + \tilde k_x \tilde k_y \partial_{\tilde R} \delta \tilde e_y}{\tilde \omega \left(\tilde \omega^2 - \tilde k_x^2 - \tilde k_y^2\right)} + i \delta \tilde e_y &=& 0
\nonumber \\
n \, \frac{\tilde R^2}{\cos \psi_\infty} \frac{\tilde k_x \tilde k_y \partial_{\tilde R} \delta  \tilde e_x + \left(\tilde \omega^2 - \tilde k_y^2\right)  \partial_{\tilde R} \delta \tilde e_y}{\tilde \omega \left(\tilde \omega^2 - \tilde k_x^2 - \tilde k_y^2\right)} - i \delta \tilde e_x &=& 0 
\label{bcfinal} \ .
\eea

%procedure to find NM
Using the fluctuation analysis techniques developed in \cite{Amado:2009ts, Kaminski:2009dh} and used in \cite{Jokela:2013hta, Jokela:2010nu}, we search for normal modes by looking for pairs $(\tilde{\omega}, \tilde{k})$ for which there is a solution to the fluctuation equations  with the boundary conditions (\ref{bcfinal}).

%%%%%%%%%%%%%%%%%%%%%%%%%%%%%%%%%%%%%%%%%%%%%%%%%%%%%%%%%%%%%%%%%%%%%%%%%%%%%%%%%%%%%%%%%%%%%%%%%%%%%%%%%

\subsection{Phonon dispersion}

In this section we analyze the spectrum of collective excitations of the anyonic superfluid with nonzero current. 
We consider a current in the $y$-direction and, as we do not expect rotational invariance, consider collective excitations with a general $(\tilde k_x,\tilde k_y)$. The mode we are most interested in is the massless phonon. We compute its dispersion as a function of the angle in the $(x,y)$-plane in which it is directed. 

\begin{figure}[!ht]
 \center
 \includegraphics[width=0.35\textwidth]{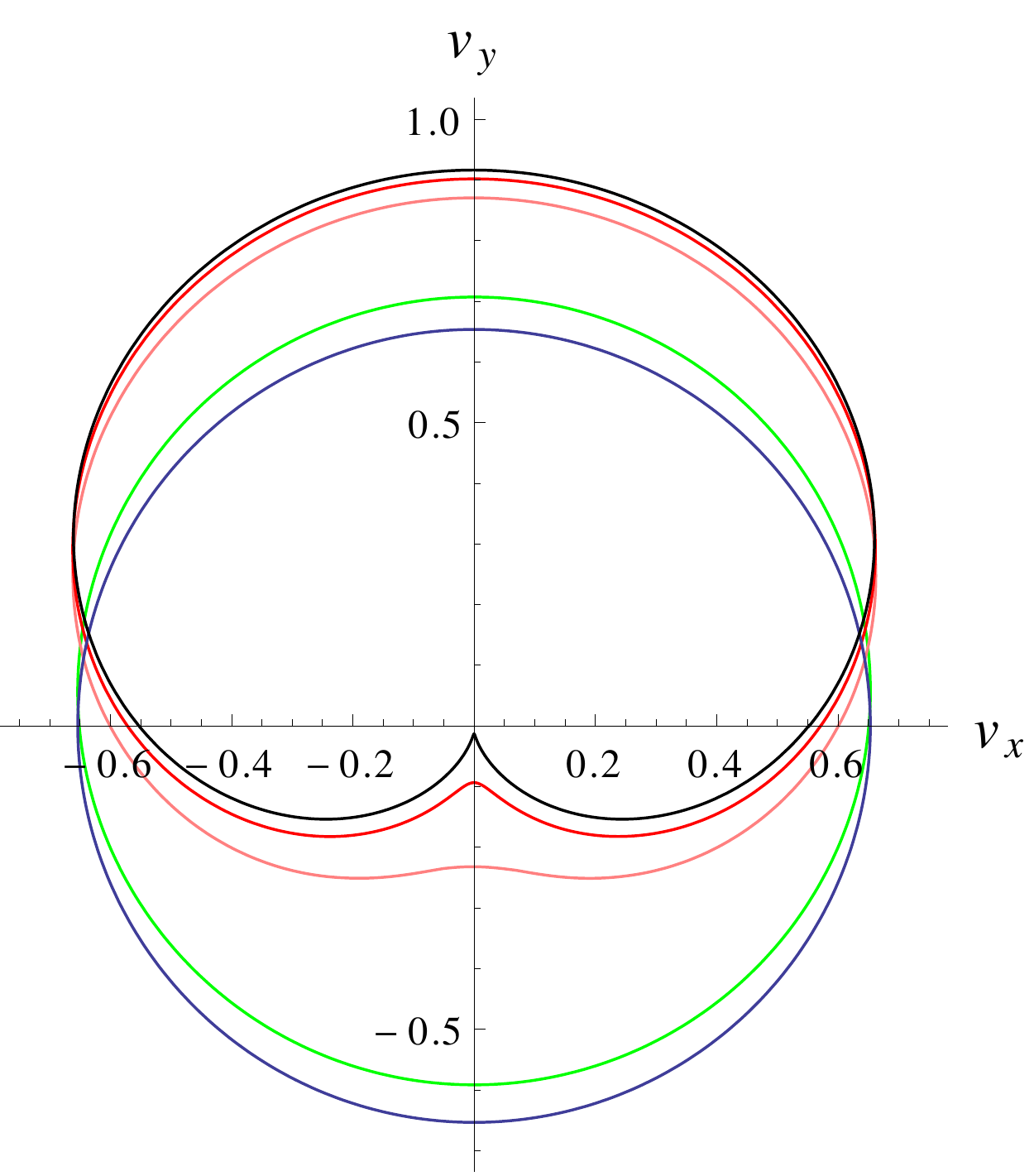}
 \includegraphics[width=0.35\textwidth]{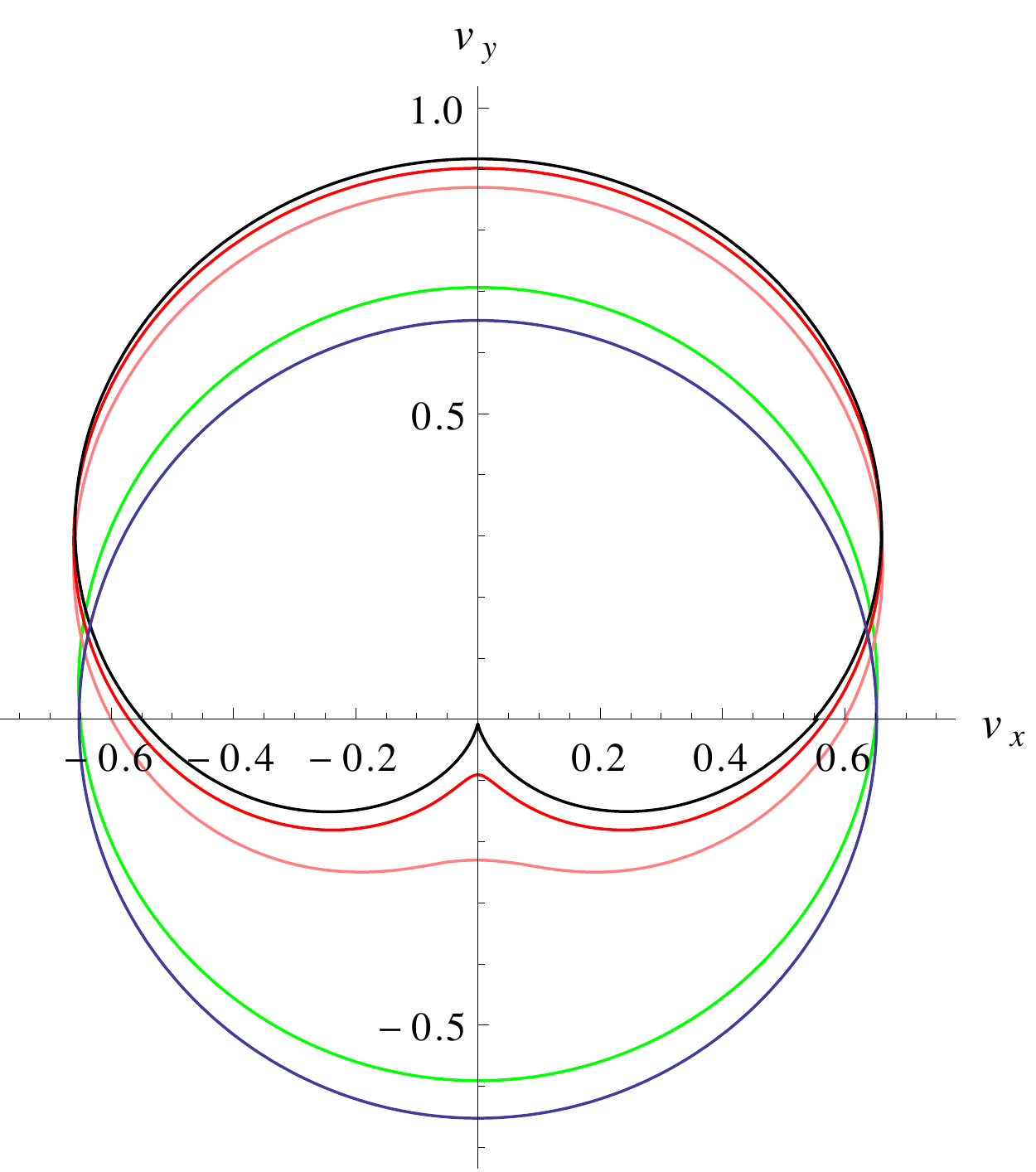} 
 \caption{The velocity of the phonon as a function of the angle for superfluid velocities $v_f = 0, 0.1, 0.5, 0.6, 0.65$ (blobs moving upwards), for $\tilde m = -8$ and for (left) $\tilde r_T=0$ and (right) $\tilde r_T=0.3$.  At $v_f \sim 0.65$, the phonon velocity in the negative $y$-direction goes to zero.}\label{fig:speedvsgamma_para_mtildem8_rTtilde0_varyingetilde}
\end{figure}

\begin{figure}[!ht]
 \center
 \includegraphics[width=0.7\textwidth]{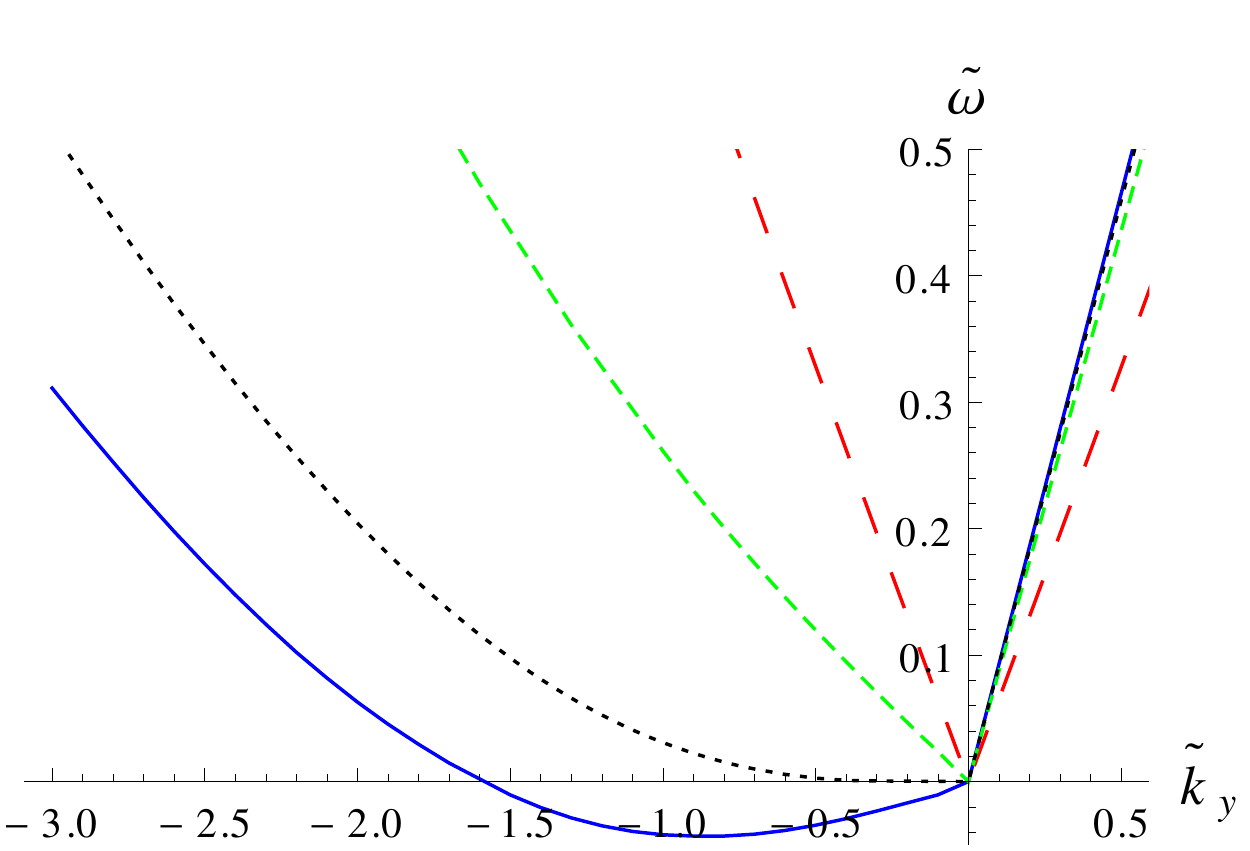} 
 \caption{The dispersion of the phonon at $\tilde r_T=0$ with  $v_f=0,0.5,0.66,$ and 0.7 for $\tilde m=-8$; for $v_f$ increasing the curves tilt counter-clockwise.  
We see that $v_{crit} =  0.66$.  The results agree with those expected from the Lorentz transformation (\ref{eq:Lorentz1})-(\ref{eq:Lorentz3}).}\label{fig:kydispersions}
\end{figure}

\begin{figure}[!ht]
 \center
 \includegraphics[width=0.8\textwidth]{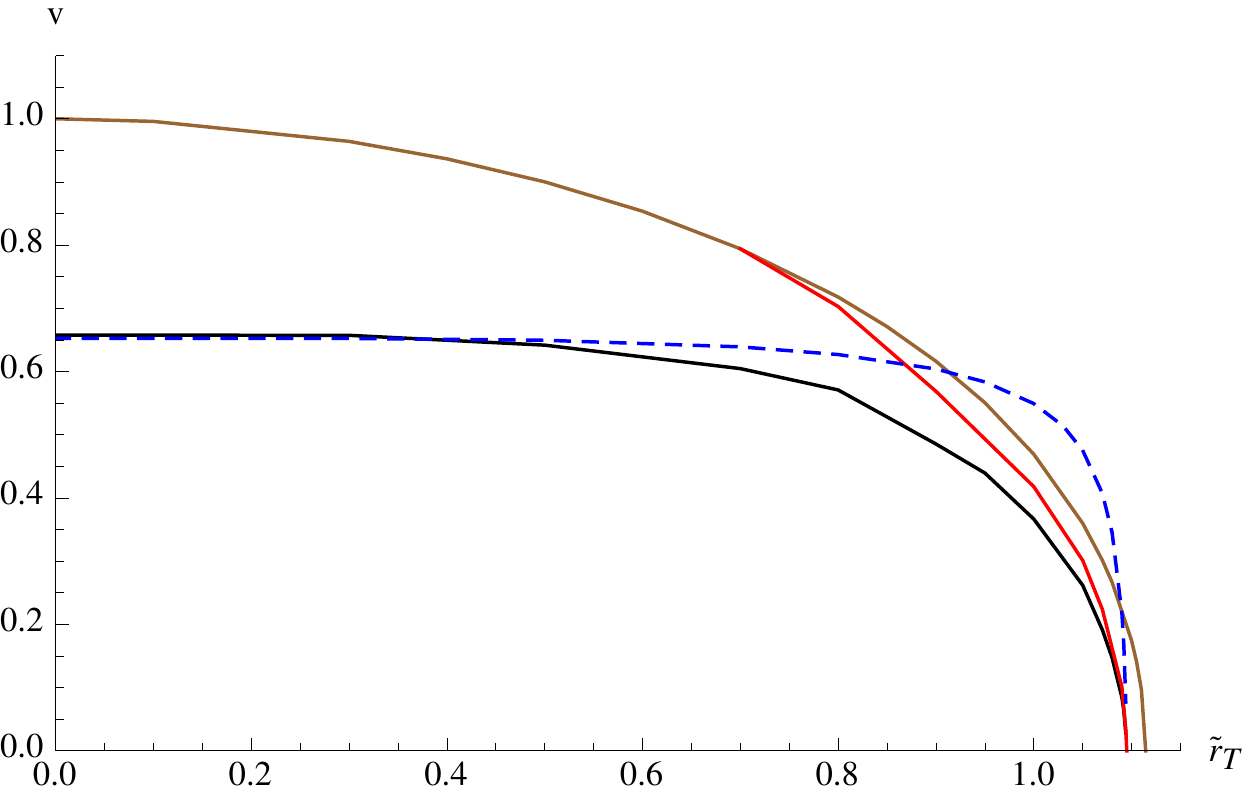} 
 \caption{The temperature dependence of various velocities: (solid curves from bottom) $v_{crit}$ (black),  $v_{complex}$ (red), and $v_{max}$ (brown). 
 In addition, the sound speed $v_s$ at zero superfluid velocity is shown as a dashed blue curve.  At $\tilde r_T=0$, $v_{crit} = v_s$ as expected from the Landau argument.  
 Also at $\tilde r_T=0$, $v_{max}=v_{complex} = 1$ which is the speed of light.  At $\tilde r_T = 1.10$, $v_{crit} = v_{complex} = v_s = 0$; above this temperature, the nonflowing superfluid is unstable.}\label{fig:ecritandvx_para_mtildem8_varyingrTtilde}
\end{figure}

When the superfluid is at rest, the phonon has an isotropic linear dispersion, 
\be
\label{lineardispersion}
\tilde \omega = v_s \tilde k
\ee
for small $\tilde k$ in any direction \cite{Jokela:2013hta}.  For $v_f >0$, the phonon velocity becomes anisotropic, as shown in Fig.~\ref{fig:speedvsgamma_para_mtildem8_rTtilde0_varyingetilde}, increasing in the direction of the superfluid flow and decreasing in the opposite direction.   Fig.~\ref{fig:kydispersions} shows the phonon dispersion in the $y$-direction for various superfluid velocities.  Similar dispersions have been found in \cite{Amado:2013aea}.  At a critical superfluid velocity $v_{crit}$, the phonon velocity vanishes, and for $v_f > v_{crit}$, the frequency of backward-directed phonons becomes negative.  We show in Fig.~\ref{fig:ecritandvx_para_mtildem8_varyingrTtilde} the temperature dependence of both $v_{crit}$ and the sound speed in the static superfluid $v_s$.

At $T=0$, the spectrum of excitations for the flowing superfluid is just given by a Lorentz transformation of the spectrum of a static superfluid. 
The nonzero current can be obtained by boosting an observer by $v_f$ in the negative $y$-direction;  the frequency and wave number of fluctuations transform as:
\bea
 \tilde\omega' & = & \gamma(\tilde\omega+v_f \tilde k_y) \label{eq:Lorentz1} \\
 \tilde k'_x   & = & \tilde k_x  \label{eq:Lorentz2} \\
 \tilde k'_y   & = & \gamma(\tilde k_y + v_f \tilde\omega) \ ,\label{eq:Lorentz3}
\eea
where the Lorentz factor $\gamma = \frac{1}{\sqrt{1-v_f^2}}$.

Plugging in the linear phonon dispersion (\ref{lineardispersion}) into (\ref{eq:Lorentz1}) gives the dispersion at nonzero $v_f$.  
In particular, the critical velocity $v_{crit}$, at which $\tilde\omega' = 0$ for negative $\tilde k_y$, is exactly $v_s$.  
To the accuracy of our numerical computations, this dispersion matches the numerical result shown in Fig.~\ref{fig:kydispersions}. 

According to the Landau criterion, $v_{crit}$ is the largest current velocity for which the anyonic fluid remains a stable superfluid.\footnote{For discussions on the Landau criterion in 
other holographic contexts, see \cite{Keranen:2010sx,Amado:2013aea}.}
Indeed, as one goes to $v_{f} > v_{crit}$, there is a negative energy mode which signals an instability towards a different configuration. 
However, we wish to emphasize that the frequency of this mode continues to be real: ${\rm{Im}} \ \tilde\omega=0$.
The remaining configuration should just be a superfluid with a lower velocity.
At zero temperature, the critical velocity for the anyonic superfluid
is found to be exactly the speed of sound when the fluid is at rest, {\emph{i.e.}} $v_{crit} = v_s$.  At nonzero 
temperature, the critical superfluid velocity is smaller than the speed of sound at rest; that is, $v_{crit}(T) < v_s(T)$. 
When one tries to give the current a velocity above $v_{crit}$, the phonon velocity becomes negative.  If the fluid passes any barrier, 
it can excite modes with a negative energy, which is just the statement that the fluid flow is no longer dissipationless.

However, if there is no barrier, the fluid flow is still stable, and the existence of the negative velocity does not make the flow unstable. 
This is clear from the zero-temperature case where the flowing superfluid is just the stable static superfluid in a boosted reference frame. 
On the other hand, at nonzero temperature, the usual description of a superfluid consists of a superfluid component and some regular fluid component. 
If this is the case, then one might expect that relative velocities between the two components could induce interactions that would excite the negative-frequency mode and make the flow unstable. Indeed, we find that at nonzero temperature, there is a velocity $v_{complex}$ at which an instability occurs.

\begin{figure}[!ht]
 \center
 \includegraphics[width=0.7\textwidth]{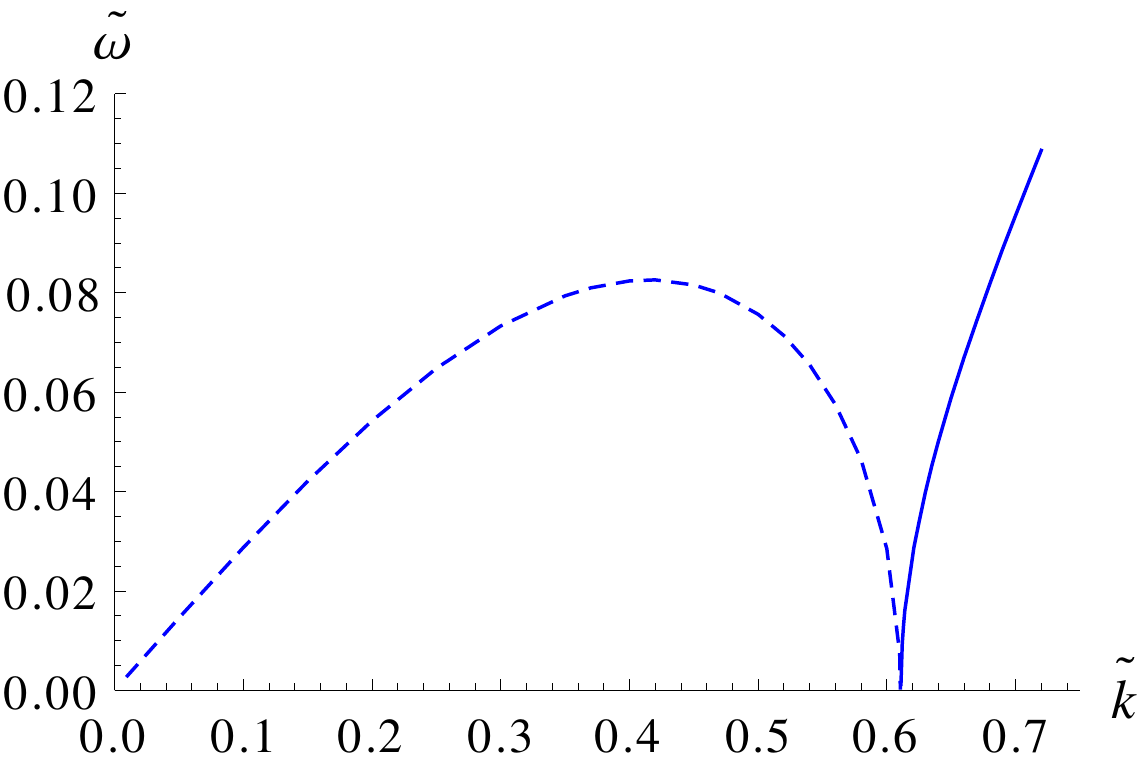} 
 \caption{The phonon dispersion for $v_f = 0$ and $\tilde r_T = 1.1$.  At this temperature $v_{crit} = v_{complex} = 0$, and consequently, $\tilde\omega$ is purely imaginary for $|\tilde k|  \lesssim 0.6$, 
 signaling an instability. The dashed curve shows ${\rm Im} \tilde \omega$.  For larger momenta, the frequency is real, denoted by the solid curve.}\label{fig:complexdispersion}
\end{figure}

In Fig. \ref{fig:complexdispersion}, we show a typical phonon dispersion corresponding to $v_f > v_{complex}$.  For excitations with small $\tilde k$, 
we find a positive imaginary frequency, signifying an instability. 
We interpret $v_{complex}$ as the velocity at which the flow becomes unstable due to interactions,\ with the normal component 
making it possible to excite the negative-frequency mode.  

The temperature dependence of $v_{complex}$ is shown in Fig.~\ref{fig:ecritandvx_para_mtildem8_varyingrTtilde}.  
In general, $v_{complex} > v_{crit}$, though the difference shrinks with temperature.  
At a sufficiently high temperature, we find $v_{complex} = v_{crit} = 0$. This is therefore the critical temperature above which the static superfluid is unstable.

At zero temperature, $v_{complex} = 1$, the speed of light.  This is in accord with our previous argument that at $T=0$, the flowing superfluid is just a static superfluid which has been Lorentz boosted.  The maximum $v_f$ obtainable by a boost is, of course, the speed of light, so the $T=0$ superfluid should be stable for any $v_f < 1$.

Note that this is a bit different than the results found in \cite{Amado:2013aea}, where the authors found that $v_{crit}=v_{complex}$. We believe this is due to the relative high temperature at which they were working, where we find the two velocities become very close. On general grounds, however, $v_{crit} \not= v_{complex}$ since at $T=0$ the dispersion is fixed by Lorentz invariance.

As discussed in Sec.~\ref{sec:backgroundnumerics}, at even higher velocities, we encounter a $v_{max}$, the maximal superfluid velocity above which no MN solution exists on the stable branch, and whose temperature dependence is also shown in Fig.~\ref{fig:ecritandvx_para_mtildem8_varyingrTtilde}. Interestingly, $v_{max}(T=0)=1$ for all values of the mass. 
The embedding geometry somehow knows that the highest superfluid velocity possible in the boundary is the speed of light.

%%%%%%%%%%%%%%%%%%%%%%%%%%%%%%%%%%%%%%%%%%%%%%%%%%%%%%%%%%%%%%%%%%%%%%%%%%%%%%%%%%%%%%%%%%%%%%%%%%%%%%%%%%%%%%%%%%%%%%%%%%%%%%%%%%%%%%%%%%%%%%%%%%%%%%%%%%%%%%%%%%%%%
%%%%%%%%%%%%%%%%%%%%%%%%%%%%%%%%%%%%%%%%%%%%%%%%%%%%%%%%%%%%%%%%%%%%%%%%%%%%%%%%%%%%%%%%%%%%%%%%%%%%%%%%%%%%%%%%%%%%%%%%%%%%%%%%%%%%%%%%%%%%%%%%%%%%%%%%%%%%%%%%%%%%%
%%%%%%%%%%%%%%%%%%%%%%%%%%%%%%%%%%%%%%%%%%%%%%%%%%%%%%%%%%%%%%%%%%%%%%%%%%%%%%%%%%%%%%%%%%%%%%%%%%%%%%%%%%%%%%%%%%%%%%%%%%%%%%%%%%%%%%%%%%%%%%%%%%%%%%%%%%%%%%%%%%%%%
%%%%%%%%%%%%%%%%%%%%%%%%%%%%%%%%%%%%%%%%%%%%%%%%%%%%%%%%%%%%%%%%%%%%%%%%%%%%%%%%%%%%%%%%%%%%%%%%%%%%%%%%%%%%%%%%%%%%%%%%%%%%%%%%%%%%%%%%%%%%%%%%%%%%%%%%%%%%%%%%%%%%%
%%%%%%%%%%%%%%%%%%%%%%%%%%%%%%%%%%%%%%%%%%%%%%%%%%%%%%%%%%%%%%%%%%%%%%%%%%%%%%%%%%%%%%%%%%%%%%%%%%%%%%%%%%%%%%%%%%%%%%%%%%%%%%%%%%%%%%%%%%%%%%%%%%%%%%%%%%%%%%%%%%%%%

\section{Discussion}
\label{sec:discussion}

% what we did: can do T=0, v_crit vs. v_complex, v_max = causality
We have presented a holographic model of a flowing, strongly-coupled anyonic superfluid.  A particularly elegant feature of this model is that, because it is based on a probe brane taking a MN embedding, there is no difficulty considering the zero-temperature limit.  By analyzing the fluctuations, we found the critical superfluid velocity $v_{crit}$ at which the phonons can begin dissipating energy and showed that at zero temperature this critical velocity was equal to the phonon velocity $v_s$, as argued by Landau.  We further found that at an even higher velocity $v_{complex}$, the superfluid is in fact unstable.

% what's left to figure out, advertisement for stripes paper and BH instability with n > 0 paper
A interesting open question is what actually happens to the anyonic superfluid when $v_f > v_{complex}$?  For $v_f > v_{crit}$, the negative energy modes, if they are excited, simply act to slow down the superfluid until it is back to the critical velocity.  However, for $v_f > v_{complex}$, the outcome is less clear.   
At sufficiently high temperature it is possible that the stable configuration is a BH embedding, corresponding to a metallic, conducting state rather than superfluid. This black hole embedding should obey the same boundary conditions as the flowing superfluid phase, which are $E^{*}=B^{*}=0$. The only such BH solutions are those with $B$ and $D$ the same as in the superfluid solution but with $E_{x}=E_{y}=0$.   
However, such solutions only exist at high enough temperature; for instance, for $\tilde{m}=-8$ such BH solutions only exist for $\tilde{r}_{T}>0.982$, so at lower temperatures at least, this can not be the end point.
Another option is that since the  unstable modes occur also at nonzero momentum, perhaps the stable ground state is spatially modulated.  An upcoming work \cite{upcoming1} will investigate more generally the instabilities of the alternatively quantized system, and in another \cite{upcoming2} we will solve for  the inhomogeneous ground state to which these instabilities lead.

\vspace{0.5cm}

{\bf \large Acknowledgments}
We thank Irene Amado, Daniel Are\'an, Andy O'Bannon, Alfonso Ramallo, Gordon Semenoff, Henk Stoof, and Stefan Vandoren for discussions.
N.J. is funded in part by the Spanish grant FPA2011-22594, by the Consolider-Ingenio 2010 Programme CPAN (CSD2007-00042), by Xunta de Galicia (GRC2013-024), and by FEDER. N.J. is
also supported by the Juan de la Cierva program.  N.J. is in part supported by the Academy of Finland grant no. 1268023.
The work of G.L was supported in part by the Israel Science Foundation under grant no.~504/13 and in part by a grant from the GIF, the German-Israeli 
Foundation for Scientific Research and Development under grant no. 1156-124.7/2011.
M.L. is supported by funding from the European Research
Council under the European Union's Seventh Framework Programme (FP7/2007-2013) /
ERC Grant agreement no.~268088-EMERGRAV.  
N.J. wishes to thank University of British Columbia for warm hospitality.
In addition, we thank the ESF Holograv Network and $\Delta$-ITP for supporting the ``Workshop on Holographic Inhomogeneities", during which this work was finalized.

\end{document}